# Multi-Wavelength transparent Microfluidics for UV-Visible spectroscopy and X-ray Scattering studies of Photoactive Systems

Benedetta Marmiroli,*[a] Sumea Klokic,*[a] Barbara Sartori, [a] Marie Reißenbüchel, [a] Alessio Turchet, [b] and Heinz Amenitsch[a]

*a Institute of Inorganic Chemistry, Graz University of Technology, Stremayrgasse 9/IV, 8010, Graz, Austria*
*b Elettra-Sincrotrone Trieste, SS 14 Km 163,5, AREA Science Park, 34149, Basovizza, (Trieste), Italy*

email: benedetta.marmiroli@tugraz.at, sumea.klokic@tugraz.at

Microfluidic devices are increasingly used in synchrotron-based experiments to deliver and probe liquid samples, offering advantages such as minimal sample consumption and reduced radiation damage. Despite their growing use, few devices have been specifically designed for monitoring liquids under photoexcitation, a promising approach for fast structural transitions. Here, a microfluidic device is presented that is transparent to X-rays in one direction and simultaneously transmits UV and visible light in the perpendicular direction. The device is fabricated using lamination and UV lithography on a dry-film resist, eliminating the need for cleanroom facilities and simplifying production. Its multi-wavelength transparency was validated through UV–Vis spectroscopy, where photoexcitation at different wavelengths induced reversible trans–to-cis isomerization of azobenzene and fluoro-azobenzene. X-ray transparency was verified through small-angle X-ray scattering (SAXS) measurements on hemoglobin and CO-ligated hemoglobin, both of which are sensitive to quaternary structural changes. These results confirm the suitability of the device for resolving protein structures and photoinduced conformational dynamics. The design also allows temperature-jump (T-jump) and time-resolved pump–probe experiments, as demonstrated by initial proof-of-concept results, providing a versatile platform for studying structural evolution in liquid samples using synchrotron SAXS.

## Introduction

Light-responsive materials have emerged as a central focus in diverse areas of the natural sciences. Upon illumination, photoactive molecules and supramolecular assemblies can undergo reversible structural changes altering properties such as geometry, polarity, or flexibility.[1] Light is a particularly attractive stimulus for controlling biological functions, as its intensity and wavelength can be precisely regulated, enabling non-invasive, localized activation with micrometer-scale spatial resolution.[2] A well-known example is the photoactive molecule retinal, which enables vision in mammals through a light-induced conformational change. This has inspired the development of artificial systems such as photoreversible DNA and protein binding.[3,4] Moreover, incorporating photoactive groups into biomolecules has become an effective strategy to modulate biological activity, as demonstrated by photoactivated drugs that exhibit light-controlled enzyme activity,[1,5] receptor affinity,[6] or targeted antibiotic action.[1] Among these photoactive groups, those that undergo photoisomerization such as azo-groups, commonly referred to as photoswitches, are among the most extensively studied.[7–10] Because light offers excellent spatiotemporal resolution, it is a highly advantageous stimulus for probing photoinduced structural dynamics. Investigating the dynamics during photoexcitation of proteins such as Rhodopsin[11] and CO-ligated hemoglobin[12] (HbCO), or of lipids specially tailored for optopharmacology (the use of light to control drug activity),[10] is crucial for understanding the timescales of light-induced processes.[13] Several structural investigations have already been conducted using techniques based on X-rays, generated by high brilliance synchrotron radiation sources, particularly X-ray Diffraction, Wide Angle X-ray Scattering (WAXS), and Small Angle X-ray Scattering (SAXS).[8,14–19] Among these, synchrotron SAXS is widely used in structural biology for nanostructural characterization typically in the 1-100 nm range. These techniques are particularly suitable for time-resolved measurements, as demonstrated by Cammarata et al. Here, time-resolved WAXS was used to reveal coordinated molecular motions during carbon monoxide (CO) binding and photodissociation in hemoglobin (Hb).[12] Such insights are essential for understanding the transition pathway from excitation to subsequent structural dynamics, and highlight the increasing use of photosensitive proteins as model systems to investigate inter- and intramolecular motions that are fundamental to biological and medical processes.[20]

However, such investigations remain challenging mainly due to the constraints imposed on suitable sample holders. In strongly absorbing samples, light penetration is limited, so and consequently the illuminated volume must be comparable to or smaller than the optical attenuation length.[21] Microfluidic channels address this issue effectively, as their geometry can be precisely designed to achieve sample depths on the order of the optical penetration length, ensuring uniform excitation of the entire sample volume. For protein samples in particular, microfluidic devices have become increasingly popular,[22,23] offering several key advantages: low sample consumption (critical for scarce protein systems), precise control of diffusion within microchannels (enabling rapid and accurate mixing), and efficient heat transfer due to their high surface-to-volume ratio, making them ideal for studying sensitive biological processes. Microfluidic devices enhance the efficiency of solution-based synchrotron X-ray experiments, particularly SAXS, offering a versatile platform for investigating light-responsive samples while simultaneously enabling precise manipulation and delivery of liquids.[24–26] Beyond these advantages, microfluidic devices connected to pumps continuously refresh the liquid sample in front of the X-ray beam, mitigating radiation damage effects provoked by its high intensity.[27] Although several microfluidic devices have been reported for light manipulation of samples, only a limited number allow multi-wavelength transparency.

In the past, a square capillary (0.1 x 0.1 mm) was connected to two reservoirs and used for manipulating starch granules with optical tweezers, while simultaneously measuring their structure with SAXS.[28,29] However, the size of the device was not scalable and was limited to microfocus beamlines, where the dimension of the X-ray beam is smaller than the dimension of the channel to avoid refraction effects due to the channel walls. A similar approach was later used to combine optical tweezers with X-ray fluorescence.[30] More recently, a microfluidic device with a sample mixer based on careful material selection for UV and X-ray transparency was reported that allows spatially separated, yet continuous SAXS and UV-Vis monitoring.[31]

In this work, both the selection of materials and the development of a suitable fabrication protocol for a microfluidic device specifically engineered for combined photoexcitation/X-ray experiments are addressed. A schematic of the device and its constituent materials is shown in Figure 1(a). To achieve transparency to X-rays along the beam path and to UV and visible light in the perpendicular direction, SUEX dry film and impact modified poly(methyl methacrylate) (IM-PMMA) were selected. The microchannel was fabricated by a sequence of lamination and optical lithography steps, with the detailed protocol outlined in this work. While SUEX has already been employed for microfluidic circuits,[32,33] to our knowledge it has never been used with SAXS. The transparency of the device to light at 365 nm, 450 nm and 513 nm was tested by photoisomerization of fluoro-azobenzene and azobenzene using UV-Vis spectroscopy, while its X-ray transparency was assessed using synchrotron SAXS measurements on Hb and HbCO protein solutions. The microfluidic channel depth was designed to remain below the UV attenuation length, based on previous photoexcitation studies in capillaries for HbCO. These results show the proof of principle for the performance of the microfluidic device, demonstrating its suitability to study a broad range of photosensitive samples using different experimental techniques.

## Experimental

### Fabrication process.

The selected geometry of the device consists of a single microchannel 1 mm wide, 40 mm long (15 mm in the X-ray accessible part) and 0.25 mm deep. The thickness of the device walls in the X-rays accessible part is 0.5 mm. The materials employed for the device and the shape of the channel are shown in the exploded scheme of Figure 1(a), while an outline of the fabrication process of the microchannel can be found in Figure 1(b).

The fabrication process is considered green as it is based on the dry film SUEX,[34] and it does not require a clean room facility. It can be described in the following steps:

a) Cutting of a 125 µm IM-PMMA sheet (Goodfellow Cambridge Ltd, UK) and drilling of two 1 mm diameter holes in order to add the tubing for feeding the channel in a later stage (vide infra, step e). The shape of the IM-PMMA was chosen to ensure an unobstructed path for the scattered photons (Figure1(b) step a).

b) Lamination of 250 µm thick SUEX photosensitive resist (micro resist technology GmbH, Germany) on IM-PMMA. The sample is put on a thin aluminium slide between two 400 µm thick spacers and laminated at 60°C for three times, in order to ensure the adhesion of the SUEX to IM-PMMA, using the rolling mill system Falcon K (BlackHole Lab, France) (Figure 1(b) step b).

c) UV lithography. Exposure of the laminated SUEX on IM-PMMA to UV light through an optical mask made of on an ink printed acetate transparent sheet. This rather simple mask was chosen as the dimensions of the channel and its features do not require high resolution. UV lithography was conducted using a Mask Aligner UV KUB-3 (Kloé, France) with a power density of 21 mW/cm² for 32 s. The post exposure bake was performed in oven (Carbolite Gero Ltd, UK) at 85°C for 40 minutes. At the end of the heating treatment, the sample was left to cool to room temperature inside the oven. The SUEX was then developed in mr-DEV 600 solution (micro resist technology GmbH) for 60 min (Figure 1(b) step c). The developing time was carefully tuned, as the mr-DEV 600 is partially dissolving IM-PMMA. The sample was then rinsed in isopropanol (Merck KGaA, Darmstadt, Germany) and dried with a nitrogen stream.

d) Sealing of the channel. A 100 µm thick SUEX was laminated on top of the channel at 50°C for two times using the same aluminium slide with spacers as before (see step b). Then UV lithography was performed without mask with a power density of 21 mW/cm² for 21 s. Post exposure bake was made at 85°C for 40 minutes in the oven where afterwards the sample was left cooling (Figure 2(b) step d).

e) Connection of the microfluidic device to the tubes for liquid transfer. First, two cubes made of PMMA containing a channel through their centre were glued to IM-PMMA in correspondence of its holes (Figure 2(b) step e). Subsequently, tubes of 800 µm external diameter were inserted into the cubes and glued using standard acrylic glue.

An adequate stainless steel sample holder was fabricated to hold the microfluidic circuit both in front of the X-rays and the laser. A picture showing the device mounted in the sample holder is shown in Figure 2.

## Test of the microfluidic device

### UV-vis measurements with light irradiation.

First, the transparency of the device materials and of the microchannel within the UV-Vis range was evaluated by UV-Vis spectroscopy (Agilent Cary 60, Agilent, US). The resulting spectra are shown in Figure S1 of supplementary information. After the material selection, the possibility to employ the microfluidic device in UV and visible spectroscopy measurements was tested using two photoswitches. The first is Azobenzene (AZB), which undergoes a reversible isomerization from *trans*-to-*cis* form using UV light (365 nm) and from *cis*-to-*trans* with blue light (450 nm). The second, 2,2',3,3',5,5',6,6'-octafluoroazobenzene (F8-AZB), allows the reversible photoinduced *trans*-to-*cis* isomerization using green light (532 nm), and the *cis*-to-*trans* with blue light (450 nm). AZB was purchased from TKI Chemicals, whilst F8-AZB was synthesized following the protocol reported in the paper of Rödl et al.[35] The microdevice was mounted on a transmission sample holder and inserted in the UV-Vis spectrophotometer. Subsequently, the photoexcitation experiments were conducted, where ethanolic AZB or F8-AZB solutions were filled into the microfluidic channel and irradiated by light. The solutions were prepared by dissolving 10 mg each of AZB or F8-AZB in 1 ml of ethanol and diluting to a 0.05 mM solution. More in detail, the irradiation of the AZB solution was accomplished with a 365 nm (UV) LED diode (EOLD-365, OSA OPTOLIGHT GmbH, Germany) of 6 mW power for 1 min (*trans*-to-*cis*), and a 450 nm (blue) LED (PL450B, OSRAM Opto Semiconductors GmbH, Germany) with a power density of 80 mW for 1 min (*cis*-to-*trans*). The F8-AZB solution was irradiated using a 532 nm (green) diode (LDS5, Thorlabs, US) of 1 mW power for 3 min (*trans*-to-*cis*) and again the blue diode (450 nm) for 1 min (*cis*-to-*trans*). The photo-excitation of the respective solutions was performed by illuminating through the IM-PMMA side of the microfluidic device. A UV-Vis spectrum was acquired after every illumination step, performing a triplicate measurement for every solution. As comparison, the same experiment was conducted with AZB and F8-AZB solutions in a standard 100 µl quartz cuvette. The curves of the same experiment have been normalized to the maximum value of the highest intensity spectrum, in order to better underline the changes, and to be able to compare the solution in the cuvette to the one in the microchannel. The resulting spectra are shown in Figure 4.

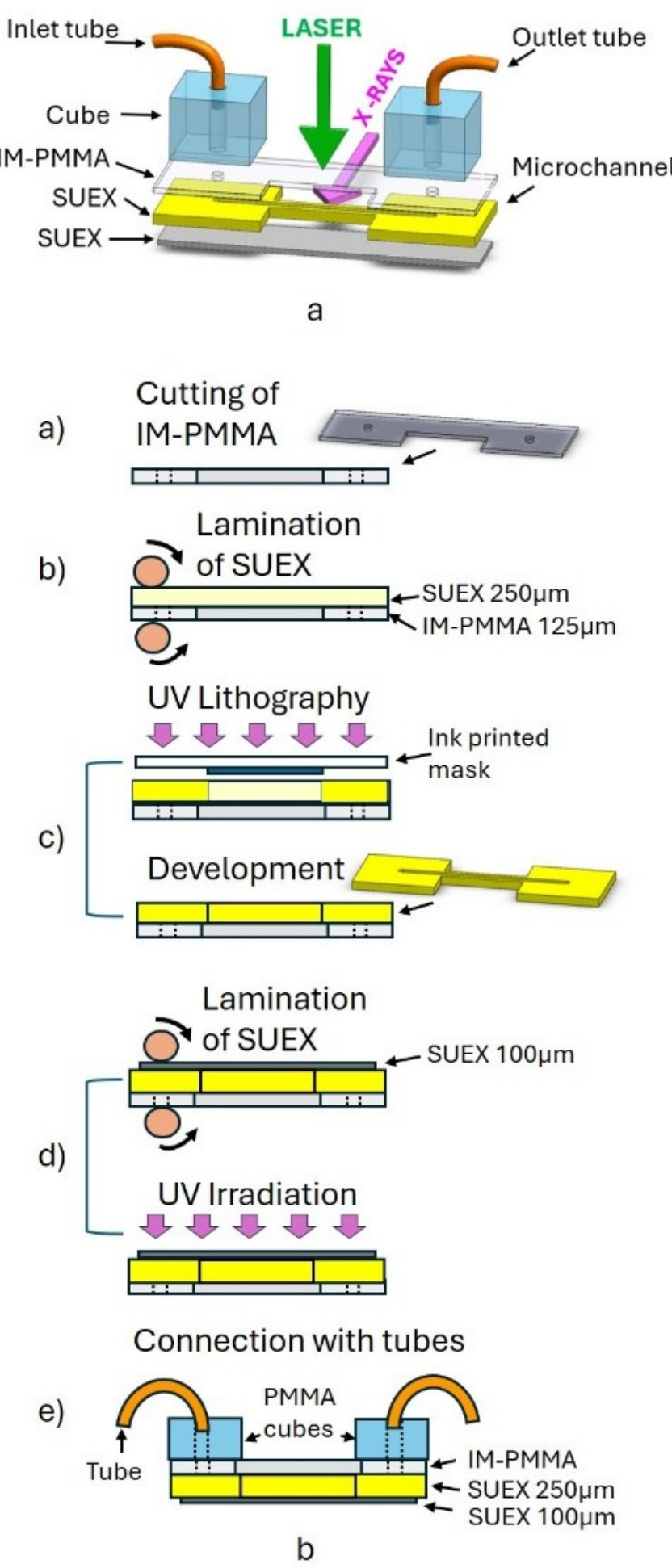

**Figure 1(a)** Exploded scheme of the microfluidic device showing the direction of the laser light and of the X-rays. The employed materials are indicated. **(b)** Scheme of the fabrication process of the proposed microchannel. **a)** Cutting of the IM-PMMA slide in the desired shape, and drilling of holes. **b)** Lamination of a 250 µm SUEX film on IM-PMMA. **c)** UV lithography through a mask and developing process of the 250 µm SUEX film. **d)** Lamination of a 100 µm SUEX film and subsequent irradiation with UV. **e)** Connection of the microchannel to external tubes.

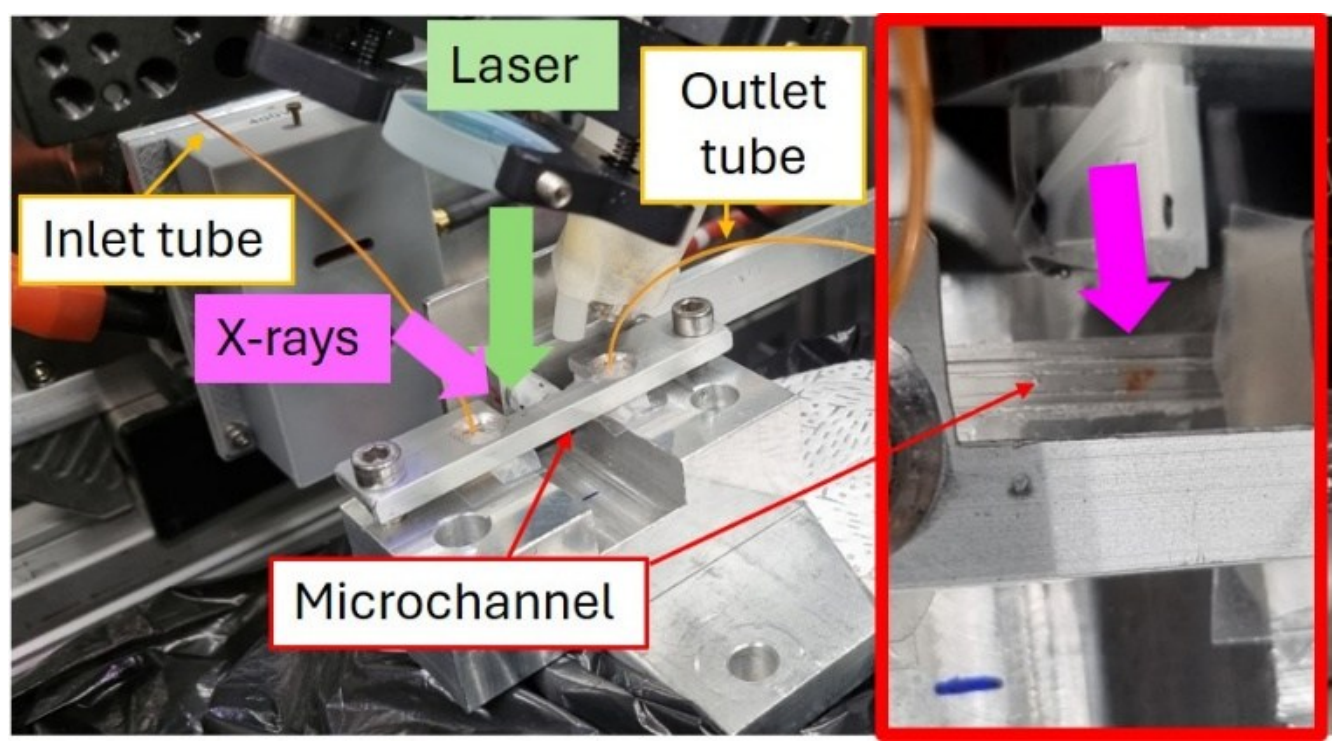

**Figure 2.** Experimental set-up of the device mounted in the sample holder at the Austrian SAXS beamline. The directions of the X-rays and laser beam are indicated by arrows, as is the rough position of the microchannel and of the tubes. On the right, in the rectangular red box, a magnification of the top view of the microchannel in the sample holder is shown.

## SAXS measurements

All X-ray scattering measurements were performed at the Austrian SAXS beamline[36] at Elettra synchrotron in Trieste (Italy). The sample-to-detector distance was set at 1250 mm providing a $q$-range $0.08 < q < 5.8$ nm$^{-1}$, where $q$ denotes the length of the scattering vector $q=4\pi/\lambda \sin(\theta/2)$, with $\lambda$ being the wavelength and $\theta$ the scattering angle. The beam size was 0.1 x 1.5 mm (vertical x horizontal). The angular scale of the detector was calibrated with silver behenate. The holder for the microfluidic channel was mounted on a motorized stage allowing to translate along the y and z direction in the plane perpendicular to the X-ray beam, and to rotate along the X-ray axis for precise alignment in transmission geometry (see Figure 3(a)). For static measurements, the SAXS patterns were acquired for 10 s, while during photo-excitation experiments 1000 images of 0.1 s were acquired. The experimental parameters for different SAXS measurements are specified in the results and discussion paragraph.

A Pilatus3 1M detector was used (Dectris Ltd, Baden, Switzerland) with active area 169 x 179 mm² and a pixel size of 172 µm, and the detector images were processed using the data pipeline SAXSDOG, developed at the Austrian SAXS beamline for automatic data reduction.[37] For data analysis, the masked full 2D pattern azimuthal integration and background subtraction of the resulting 1D data was conducted. The integrated data were processed with IGOR pro (Wavemetrics, Inc., Lake Oswego, OR). The low $q$-range part (0.38 - 0.47 nm$^{-1}$) of the scattering curves of the protein after buffer subtraction was considered to obtain the radius of gyration $R_g$,[38] which is related to the dimension of the protein, according to the equation (1) reported in paragraph 1 of supplementary information. The upper limit of the $q$-range for Guinier approximation was selected following the guidelines for the publication of SAXS data of biomolecules in solution.[39] The pair distance distribution function, whose Fourier transformation averaged over all directions in space gives the scattering intensity $I(q)$, defined by equation (2) in the supplementary information,[38] was determined using the GNOM program from the ATSAS package 4.1.1.[40] The invariant Q, second moment of the scattering curve which is related to the electron density of the sample, was determined following equation (3), while the correlation length $l_c$, related to characteristic dimensions of the sample, was calculated using equation (4) as shown in the Supplementary Information.[38]

Photoexcitation experiments performed in combination with time-resolved SAXS measurements were accomplished using the optical on-line table available at the Austrian SAXS beamline.[41] A schematic of the experimental setup is shown in Figure 3(a), emphasizing the 513 nm laser light transferred via motorized optical mirrors and lenses to overlap spatially with the X-ray beam, along with the X-ray detector and the microfluidic circuit connected to a syringe pump. The microfluidic channel was positioned on the sample holder with its UV-transparent IM-PMMA side facing the incoming laser. [41] The light to initiate the photo-switch was delivered by the Yb:KGW femtosecond laser (PHAROS, Light Conversion, Lithuania), which allows to precisely control the power and thus the photoresponse of the sample. Photoexcitation experiments were conducted at three different laser powers by adjusting the laser transmission T: 2.6 mW (T=5%), 62 mW (T=50%), and 182 mW (T=100%). The laser spot size was set to 3.5 mm FWHM.

SAXS measurements were conducted on human hemoglobin (Hb). To evaluate the scattering signal of the protein in the microchannel when compared to a standard capillary, a protein solution and the related buffer were measured for 10 s using either a quartz capillary of 1.5 mm of diameter, or the microfluidic device connected to a syringe pump (TSE System GmbH, Germany) to infuse the sample at a flow rate of 0.25 µl/min. As proof of principle of the performance of the microfluidic device under photoexcitation conditions, SAXS measurements were performed to study the structure of Hb stabilized by the presence of the ligand CO (HbCO) before and after green laser irradiation following previous literature protocols. [12]

## Preparation of hemoglobin for SAXS measurements.

25 mg of Hb (Merck KGaA, Germany) were dissolved in 1 ml of 50 mM Tris at pH 7.4. HbCO solution was prepared according to Cho et al.,[42] as follows: 25 mg of human Hb were dissolved in 10 ml of 50 mM Tris, pH 7.4. Subsequently, 10 µl of a solution of KCN 1 M were added to the Hb to obtain cyanomet hemoglobin (HbCN). Subsequently, 50 µl of 0.9 M $Na_2S_2O_4$ were added to the HbCN solution. The obtained $HbCNS_2$ solution was purged with CO gas for 30 minutes yielding HbCO.

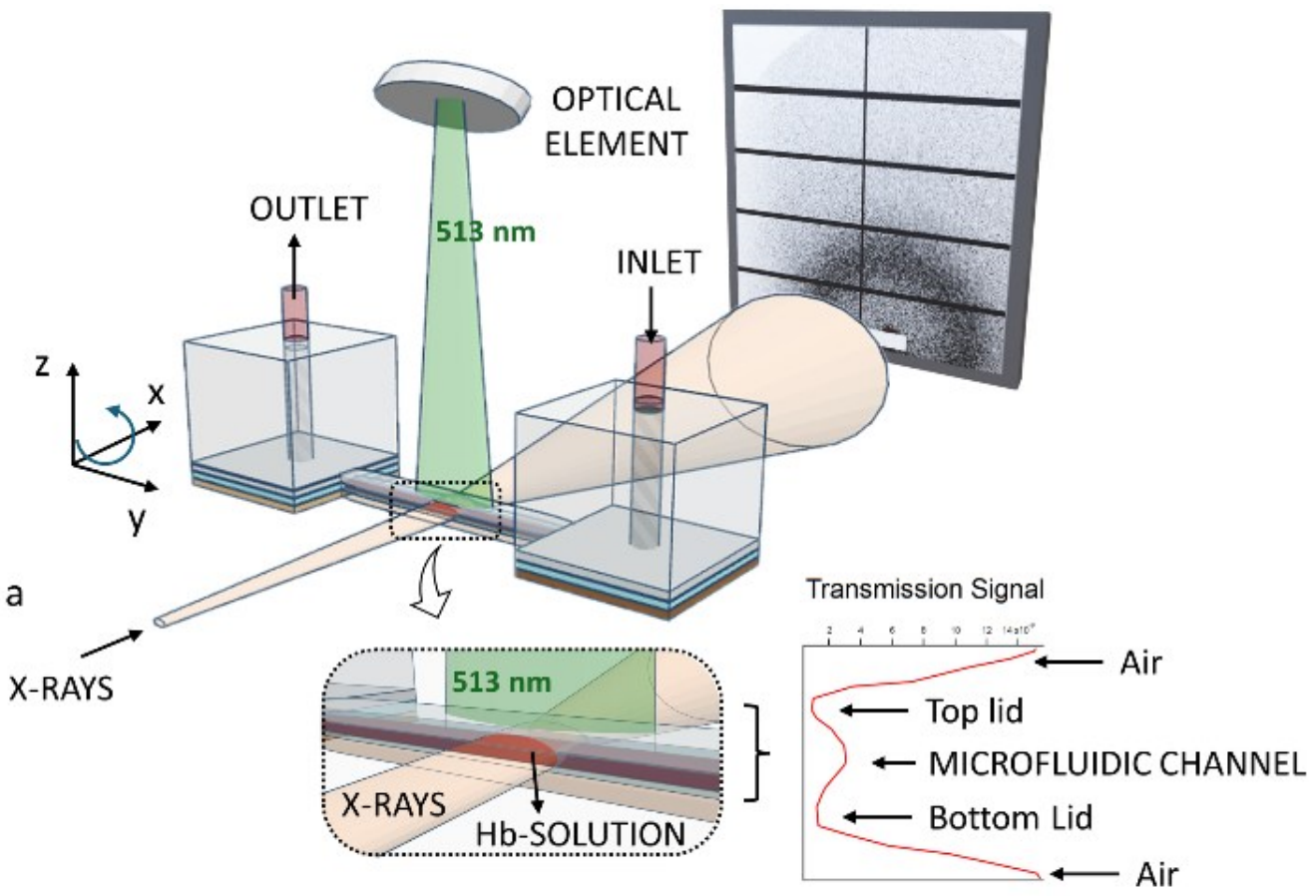

**Figure 3 (a**) Scheme of the experimental set-up showing the microchannel, the detector, the X-ray beam path and the 513 nm laser light path provided by. The direction of X-rays and of the sample movements are also shown, where x is the direction of X-rays, z is the direction of the laser perpendicular to the microchannel, y is the direction along the channel. **(b)** Magnification of the investigated area, and transmission signal employed to align the position of the channel to the X-ray beam.

## Results and discussion

### Characterization of the selected materials.

Measuring light-responsive samples, which requires both a light source and X-rays, demands a careful selection of materials for the microfluidic device, considering also the limitations due to the fabrication processes at the micro scale. In the configuration shown in Figure 3(a), where the two light sources are arranged perpendicularly, the channel walls must be both resistant and transparent to X-rays, while providing a low background scattering signal. Moreover, the top lid must withstand and be transparent to wavelengths above 340 nm, while the bottom lid only needs to withstand laser irradiation. To meet these requirements, SUEX and IM-PMMA were chosen as suitable materials to compose the device.

The transparency of the materials in the UV or visible range was evaluated by UV-Vis spectroscopy, as shown in Figure S1 of the supplementary information. For the top lid, the IM-PMMA slide of 125 µm thickness was chosen, as it is transparent down to a wavelength of 340 nm, while the bottom lid made of SUEX dry film photoresist is optically transparent. The transparency of the fabricated microchannel is also shown.

As far as the X-rays transparency and scattering are concerned, the device has been tailored for the Austrian SAXS beamline. The beamline operates at 8 keV (X-ray wavelength 0.154 nm) with a photon flux density at sample of 1012 ph.s$^{-1}$.mm$^{-2}$ at 2 GeV. The X-ray transparency of SUEX was evaluated by considering its major component (mixture of epoxy resins) and calculating the attenuation length, which is the depth into the material along the surface normal for which the X-rays intensity decays to 1/e of its value at the surface. The procedure and the details are described in chapter 2 of the supplementary information. In the present case, the attenuation length varies from 1.58 mm to 1.64 mm. Based on these values, it is possible to calculate the transmission of 0.154 nm wavelength X-rays through 0.5 mm of SUEX, which is approximately 73%. Therefore, the material is sufficiently transparent. To quantify the scattering signal of SUEX, its absolute scattering probability was calculated measuring the scattering curve of a 0.5 mm thick SUEX film and of a calibrated glassy carbon slide of 1.05 mm, following the method already reported by our group.[43] As the scattering probability of the glassy carbon, and the thickness of both glassy carbon and of SUEX are known, the absolute scattering probability can be derived from the measured scattering intensity curves. The result, presented in Figure S2, shows that SUEX has an absolute scattering probability which is comparable to 1 mm of water at q values around 1 nm$^{-1}$, i.e. 0.0016 (the scattering intensity of water is 1.6328*10$^{-2}$ cm$^{-1}$ at 293K).[44]

To ensure a measurable scattering signal of the sample within the microfluidic device, the channel dimensions were chosen to achieve sufficient sample volume. The depth of 0.25 mm is larger than the vertical dimension of the scattering X-ray beam (0.1 mm) so to measure only the center of the channel and to avoid refraction or parasitic scattering effects due to the top or the bottom channel lid. The length of the central part of the microfluidic device, accessible to X-rays, is 15 mm to allow scanning in the y direction (see Figure 3). This could be relevant for time-resolved experiments under continuous flow, since every position corresponds to a different residence time of the sample inside the channel.

### Photoexcitation performance of AZB derivatives.

To demonstrate the feasibility of photoexciting a sample in the microchannel using light in the UV and visible regime, AZB and 8F-AZB were employed. Upon irradiation, both AZB and 8F-AZB undergo reversible *trans*-to-*cis* photoisomerization (Figure 4(a) and (d)), resulting in distinct absorbance changes that can be monitored by UV-Vis spectroscopy. For this purpose, ethanolic AZB or 8F-AZB were introduced into the microfluidic channel and illuminated with 365 nm/450 nm and 532 nm/450 nm light, respectively. The corresponding normalized UV-Vis spectra before and after photoexcitation are presented in Figure 4(b) and 4(e). To ensure that the optical response of the device could be directly compared to a standard measurement setup, identical experiments were conducted in a standard cuvette, with the resulting spectra provided in Figure 4(c) and (f). As far as the photo-switching of AZB is concerned (Figure 4 (b) and (c)), the monitored absorption band corresponds to the n→π* transition which is particularly sensitive to the cis-isomer.

Irradiation by 365 nm promotes the conversion of AZB to its *cis* conformer (Figure 4(a)),[7] as evidenced by an increased absorption intensity both for the sample contained in the cuvette and for the one in the microchannel. Notably, the increase in absorbance (ΔA) observed for AZB in the microchannel (ΔA = 0.07) is more than double than that observed in the cuvette (ΔA = 0.029) under identical irradiation conditions. A similar trend was observed for 8F-AZB (Figure 4 (e) and (f)), where irradiation at 532 nm enhanced the *cis*-isomer population (Figure 4(d)) as evidenced by the increase in absorbance, while the 450 nm light (*cis*-to-*trans*) induces a shift of the absorbance band.[35] In the non-irradiated state, both *trans* and *cis* conformers coexist as clearly envisioned for 8F-AB (red traces).The microchannel exhibited a significantly larger absorbance change (ΔA = 0.36) compared to the cuvette (ΔA = 0.16). After 450 nm illumination, the 8F-AZB in the channel undergoes a shift in the absorption band of 9 nm, compared with only a 4 nm shift in the cuvette. This enhanced response arises from the much smaller photoexcited volume (1.25 µL) and optical path length (0.25 mm) of the microfluidic device, compared to the larger volume (100 µL) and 1 mm path length of the cuvette. In the cuvette, only a limited fraction of the liquid is effectively photoexcited, whereas the microchannel design maximizes optical penetration and excitation efficiency. These results confirm that the microfluidic device enables efficient photoexcitation using both UV and visible light, demonstrating effective light penetration and optical performance. Furthermore, these proof-of-concept experiments establish the suitability of the device for standard UV–Vis spectrophotometric measurements.

### SAXS measurements

Due to their sub-nanometer molecular dimensions and inherently low scattering contrast relative to the solvent, AZB and 8F-AZB are not well suited to be analysed by SAXS. Therefore, to demonstrate that the device can be used for X-ray techniques during laser-induced photoexcitation, Hb was selected as a model system because its CO-bound form (HbCO) is photosensitive and undergoes well-characterized structural changes upon photoactivation, which can be monitored by SAXS. Hb is a tetrameric protein composed of two identical dimers, and it exists in at least two different quaternary structures in solution: the relaxed (R) structure and a tense (T) state. The R structure is stabilized by adding ligands like CO, while the T state predominates when the protein is unligated. The R-to-T transition involves conformational changes both of tertiary and of quaternary structure.[45] A laser pulse in the green wavelength range can photolyze HbCO triggering the R-to-T transition.[12]

Initially, Hb was measured in a standard 1.5 mm diameter quartz capillary, commonly used for SAXS experiments, which served as a reference for subsequent measurements performed in the microchannel device. Five 10 s exposures were acquired, and the corresponding scattering curves were averaged. The SAXS pattern with the Guinier fit and the $R_g$ value are shown in Figure 5(a). After the capillary measurements, the empty microfluidic channel was positioned for SAXS experiments. Its alignment to the X-ray beam was achieved by translating the device along the z axis while monitoring

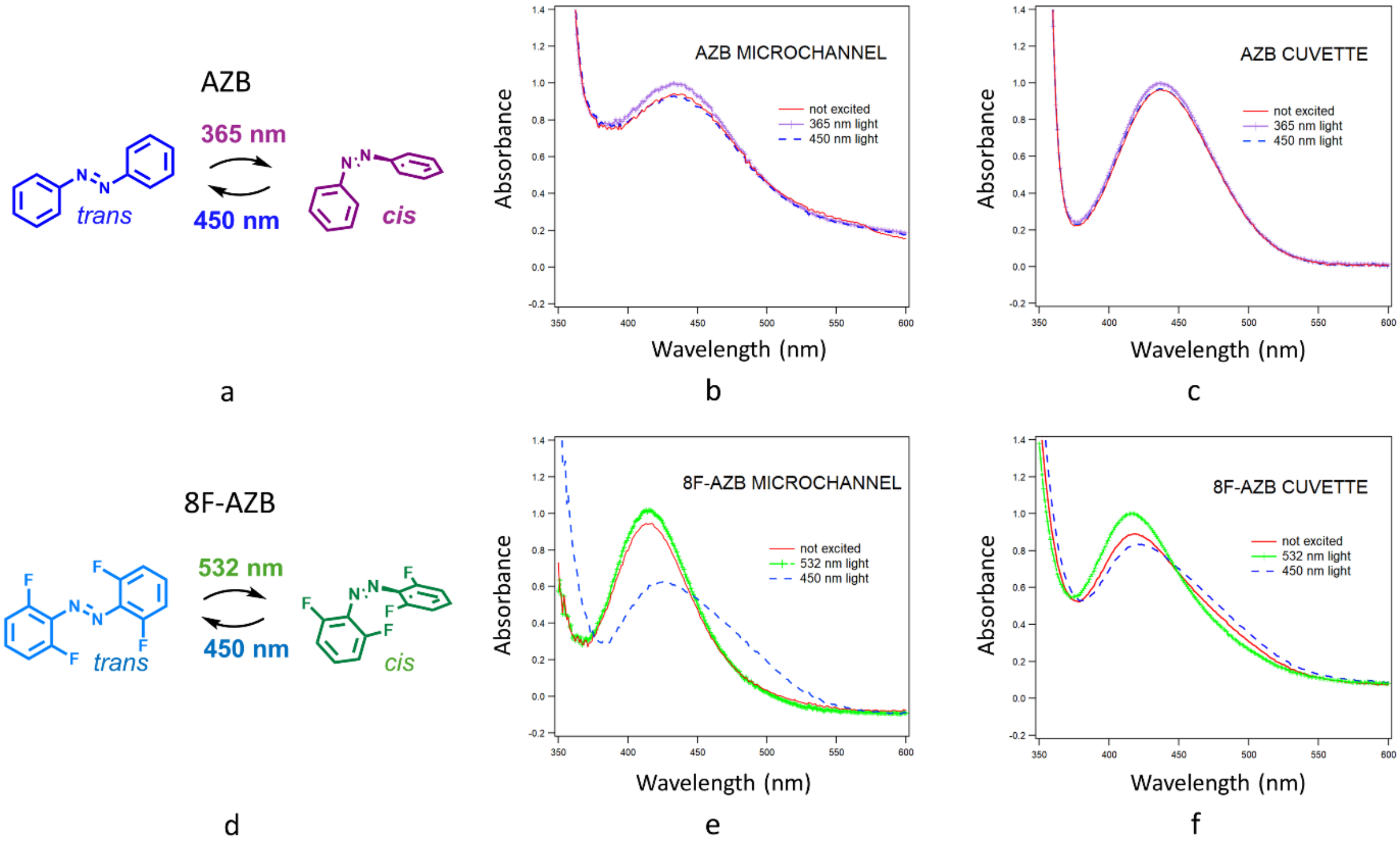

**Figure 4 (a)** Scheme of the conformational changes of AZB after photoswitching. **(b)** UV-Vis measurement of the photoswitching of AZB molecule in the microfluidic channel. Red curve is the initial AZB, violet curve with vertical bars is the AZB after 365 nm light irradiation, dashed blue curve corresponds to AZB after 450 nm light illumination to switch back. **(c)** UV-Vis measurement of AZB switching in the cuvette. **(d)** Scheme of the conformational changes of 8F-AZB after photoswitching. **(e)** UV-Vis measurement of the switching of the 8F-AZB molecule in the microfluidic channel. Red curve is the initial 8F-AZB, green curve with vertical bars is the 8F-AZB after 532 nm light irradiation, dashed blue curve corresponds to 8F-AZB after 450 nm light illumination. **(f)** UV-Vis measurement of 8F-AZB of switching in the cuvette the transmission signal with a photodiode positioned on the beamstop. The variation in transmission through the device (top lid, channel walls, bottom lid) allowed the precise identification of the channel center, as shown in Figure 3(b). Because the top and bottom lids absorb X-rays more than the channel walls the center of the channel is therefore the one in which the transmission is relatively higher.

After alignment, Hb and HbCO were measured while slowly flowing in the channel at 0.25 µl/min. Twenty 10 s exposures were collected, and the average of the corresponding scattering curves with the Guinier fit are shown in Figure 5(b). Finally, 40 exposures of 0.1 s each were taken for HbCO and averaged (result shown in Figure 5(c) together with the Guinier fit and $R_g$ value). A high protein concentration was intentionally used for both Hb and HbCO to ensure a strong and reliable scattering signal within the microchannel. Despite this, the internal volume of the device is small (10 µl) thus rendering such measurements efficient and economical, even with costly or limited protein samples. In principle, a dilution series could be performed for each protein to identify the optimal concentration for SAXS. For the concentration tested here, aggregation was observed as indicated by the strong increase in scattering intensity at $q < 0.2$ nm$^{-1}$ visible in Figure 5. Nevertheless, this does not hinder comparison between the capillary and microchannel measurements. As shown in Figure 5(a) and (b), the SAXS measurement of Hb in the microchannel is comparable to that collected in the standard capillary, demonstrating that the microfluidic device can be used to perform SAXS experiments under conditions equivalent to conventional setups. For both Hb and HbCO, fits were obtained using the GNOM program,[40] highlighting the ability of the device to distinguish subtle conformational changes in proteins. This is also evident in the corresponding pair distance distribution functions presented in Figure S3. By averaging a sufficient number of frames, the resulting scattering curves present a good signal-to-noise even in the higher $q$ range ($q = 1 - 4$ nm$^{-1}$), allowing the detection of the difference between Hb and HbCO (Figure 5(b)), which arise from the tertiary and quaternary structural changes.[12]

All curves were fitted in the same $q$-range as for the capillary SAXS data to determine the radius of gyration $R_g$. The resulting $R_g$ values shown in Figure 5 were found consistent across all measurements within the experimental error, and in agreement with previously reported literature values,[46] demonstrating that the fabricated microchannel allows SAXS measurements with short acquisition times while maintaining a good signal-to-noise ratio.

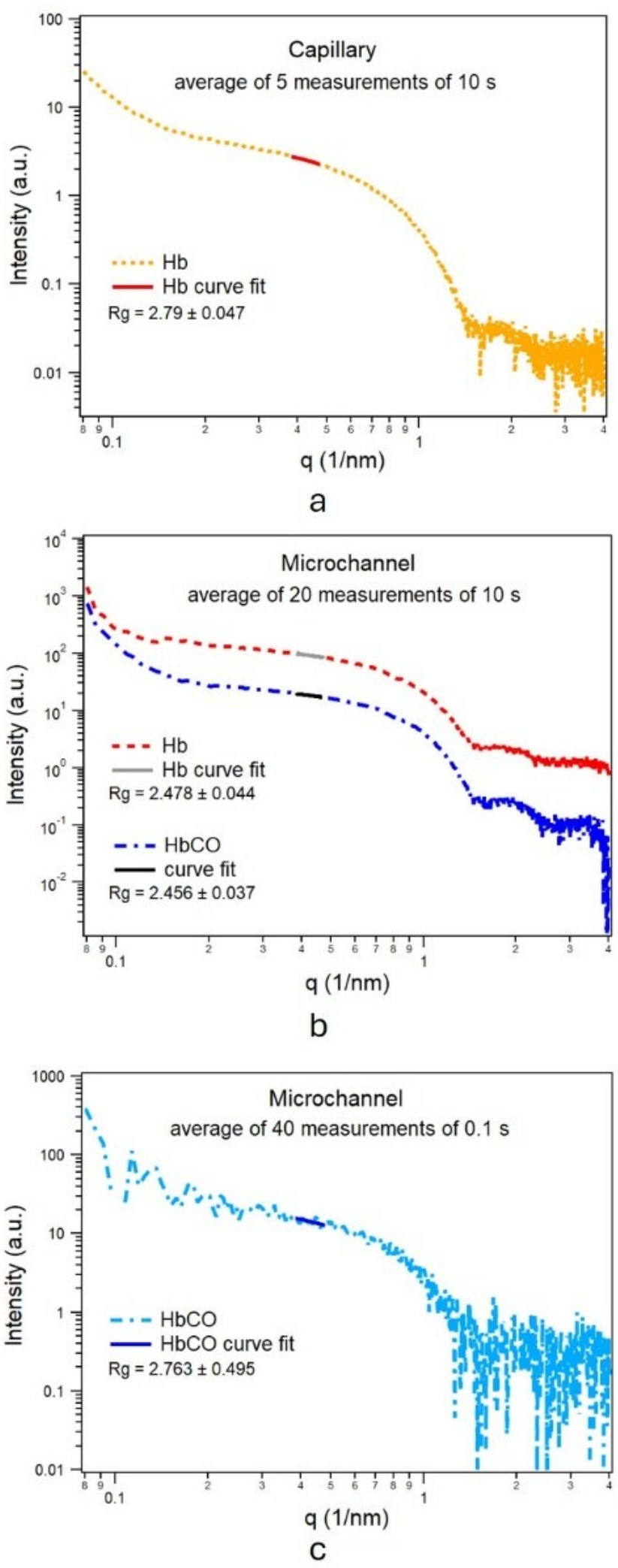

**Figure 1**. Scattering curves with corresponding Guinier fit. The values of $R_g$ are reported. **(a)** Hb in the capillary (dashed line) and fit (red continuous line). The curve is obtained averaging five measurements of 10 s. **(b)** Hb (red dashed line) with corresponding fit (continuous grey line) and HbCO (blue dashed with dots line) with corresponding fit (continuous black line) in the microchannel. The scattering curves are obtained averaging 20 measurements of 10 s. The curves are shifted in vertical for clarity. **(c)** HbCO (light blue dashed with dots line) with corresponding fit (continuous blue line) in the microchannel. The scattering curve is an average of 40 measurements of 0.1 s.

To evaluate device performance under photoexcitation conditions (513 nm), SAXS measurements for HbCO were first acquired without illumination (laser OFF), followed by measurements under continuous irradiation (laser ON). The scattering curves were compared. Three laser powers were employed to evaluate the effect of photodamage of the protein sample. For laser powers of 2.6 mW and 62 mW, the difference curves (ON minus OFF) oscillate fluctuate around zero (Figure S4, supplementary information), which indicates no photodamage. In contrast, at 182 mW, significant sample damage occurred under static conditions as well as when flowing the sample at 100 µL/min, which corresponds to a residence time of about 0.5 s under irradiation. Photoexcitation measurements on HbCO were conducted using linearly polarized light, which allowed the observation of photodamage. In contrast, Cammarata et al. used circularly polarized light inducing structural changes in the protein due to photolysis.[12]

Protein damage typically leads to adhesion to the container walls, whether in capillaries or microchannels. Hence, the microchannel was routinely cleaned after every protein measurement by flowing a 1% v/v aqueous solution of Hellmanex III, followed by Milli-Q water. Owing to the chemical resistance of SUEX, the device can be efficiently cleaned and reused multiple times, as evidenced by identical buffer scattering curves shown in Figure S6 before and after cleaning.
In the next step, experiments were conducted by continuously acquiring SAXS images with 0.1 s exposure time first with the light off, and then switching it on to monitor the changes. Again, three laser power values were tested (P= 2.6 mW, 62 mW and 182 mW), with HbCO and buffer measured under a flow rate of 100 µl/min. For completeness, the empty microchannel was also measured. From each dataset, the invariant and the correlation length were determined as a function of time (Figure 6 and Figure S7). At 2.6 mW and 62 mW, the correlation length remained constant for protein, buffer and empty device, whereas at 182 mW, changes were observed only for the protein. As the correlation length is related to the characteristic dimensions of the protein, these results confirm the absence of radiation damage at P = 2.6 and 62 mW, while at P= 182 mW protein degradation occurs due to photodamage.
When considering the invariant, no changes are observed at 2.6 mW, whilst at 62 mW and 182 mW, illumination produced a pronounced signal jump, that increased with laser power. This effect was observed for the protein, the buffer, and the empty channel. This behavior is attributed to a laser-induced thermocapillary effect,[27] where energy from the laser heats the microchannel and the sample, resulting in a thermal signal which then overlaps with the scattering signature of the protein structural changes. A similar effect was reported by Cammarata et al.,[12] who suggested decoupling heating and photolysis contributions by subtracting the signal taken after photolysis recovery but before thermal relaxation (for HbCO, this occurs within 32 ms). The magnitude of the thermal jump increases with laser power, suggesting that the microchannel could, in principle, be used for temperature-jump (T-jump) experiments. In such applications, the contribution of the channel itself must be considered, and the T-jump can be calibrated by measuring samples with known structural transitions at defined temperatures. This approach has been previously demonstrated by our group using a phospholipid for slow temperature changes,[47] and by Prassl et al. using lipoproteins for fast variations.[48]
These preliminary SAXS experiments demonstrate that the described microfluidic device can be used for photoexcitation and T-jump studies on proteins, as well as on other photosensitive samples, highlighting its versatility and potential for a wide range of time-resolved structural investigations.

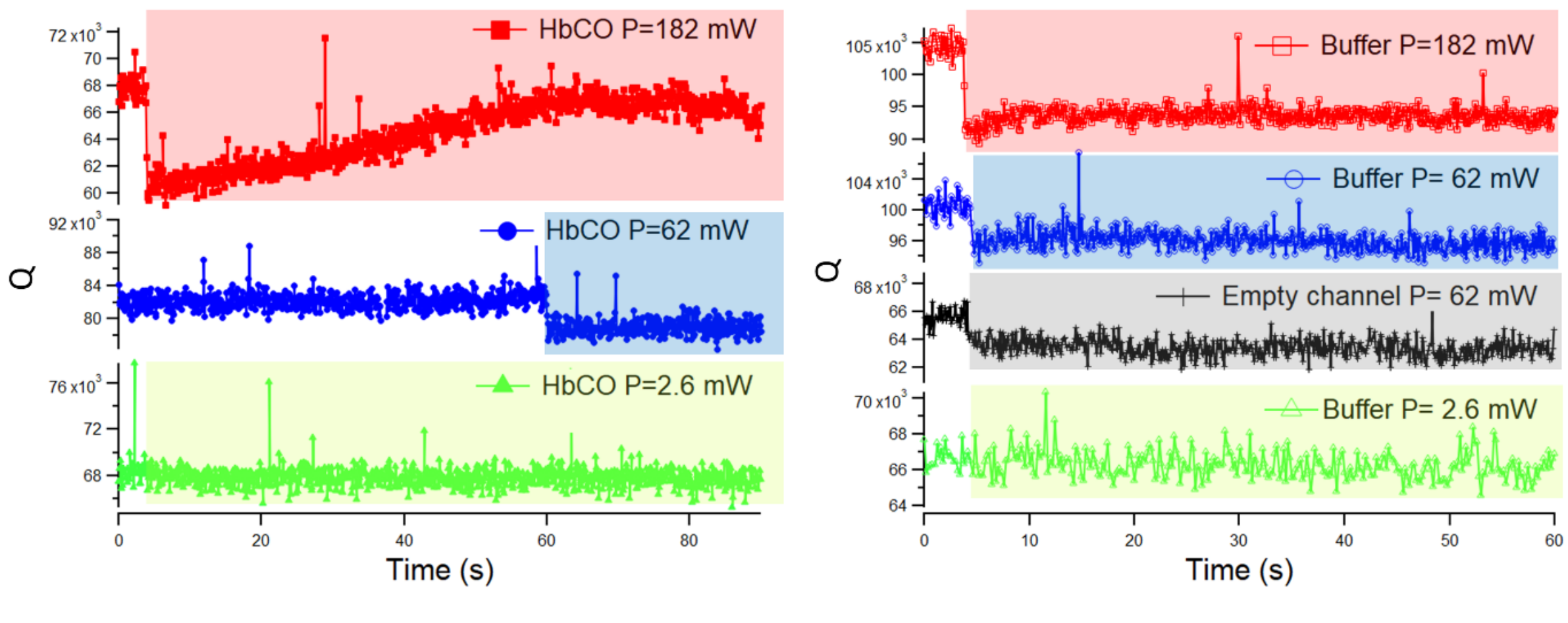

a b

**Figure 2 (a)** Calculated invariant *vs* SAXS exposure time for the protein inside the microfluidic device flowing at 100 µl/min. During the SAXS measurement the laser is turned ON. The shadowed area corresponds to the measurements done with laser ON. Different laser power values are shown. The laser with P=62 mW was turned on around one minute after starting the measurement. **(b)** Calculated invariant vs SAXS exposure time for the buffer inside the microfluidic device flowing at 100 µl/min and for the empty channel.

## Conclusions

In this work, a microfluidic device is developed that is transparent to visible and UV light down to 340 nm in one direction, and to X-rays in the perpendicular direction. Fabricated using standard UV lithography combined with SUEX dry resist lamination on IM-PMMA, the device does not necessarily require a cleanroom and can be downscaled to tens of microns, limited only by the dry film resist resolution. The device was successfully employed for photoexcitation experiments monitored by either UV-Vis spectroscopy or SAXS. Its small channel dimensions ensure full illumination of the sample across the investigated wavelengths (365 nm, 450 nm, 513 nm). SAXS measurements of proteins in the microchannel demonstrate that subtle structural differences, such as between Hb and HbCO, can be detected while fully exploiting the low sample consumption allowed by microfluidics. Generally, the device dimensions can be tailored to match both the adsorption length of the excitation light, and the size and penetration depth of the X-ray beam, thus optimizing the photoexcitation conditions for SAXS experiments. The device is easily cleaned and reused, as its materials are compatible with standard protein and lipid cleaning solutions. This versatile platform enables photoexcitation studies and laser-induced thermal jump (T-jump) experiments. While its performance has been demonstrated on proteins, it can be readily applied to inorganic or hybrid systems, or in experiments that require simultaneous optical excitation and X-ray probing (pump-probe configurations). More complex device geometries are also possible, including mixing regions before the observation channel. Moreover, the device could be operated in a 'light-reconfigurable' way[27] where injection, pumping, mixing, and sorting are controlled by light in one direction, while structural characterization is concurrently performed with X-rays in the perpendicular direction. By combining microfluidics with simultaneous optical and X-ray transparency, this platform overcomes limitations of illumination efficiency, detection of subtle structural changes and temporal control, enabling new opportunities for time-resolved structural studies across biology, chemistry, and materials science.

## Author contributions

BM conceptualization, investigation, device fabrication, formal analysis, draft writing. S.K. conceptualization, investigation, formal analysis. B.S investigation, formal analysis. A.T and M.R device fabrication. A.H. Conceptualization, SAXS investigation, formal analysis, fund raising. All authors contributed to review and editing of the manuscript.

## Acknowledgements

The authors gratefully thank A. Radeticchio and C. Morello for technical assistance, and Prof. H. A. Reichl and M. Rödl for providing the fluoro-azobenzene compounds. The authors further acknowledge Y. Roque-Diaz and Dr. G. Milcovich for some helpful assistance to experiment (CERIC-ERIC proposal no. 20237222)

## Notes and references

1 A. W. A. Velema, J. P. Van Der Berg, M. J. Hansen, W. Szymanski, A. J. M. Driessen and B. L. Feringa, *Nat Chem*, 2013, **5**, 924–928.
2 Z. Chu, C. A. Dreiss and Y. Feng, *Chem Soc Rev*, 2013, **42**, 7174–7203.
3 G. Mayer and A. Hechel, 2006, *Angew Chem Int Ed*, 2006, **45**, 4900-4921.
4 J. Andersson, S. Li, P. Lincoln and J. Andréasson, *J Am Chem Soc,* 2008, **130**, 11836–11837.
5 M. Borowiak, W. Nahaboo, M. Reynders, K. Nekolla, P. Jalinot, J. Hasserodt, M. Rehberg, M. Delattre, S. Zahler, A. Vollmar, D. Trauner and O. Thorn-Seshold, *Cell*, 2015, **162**, 403–411.
6 I. Tochitsky, M. R. Banghart, A. Mourot, J. Z. Yao, B. Gaub, R. H. Kramer and D. Trauner, *Nat Chem*, 2012, **4**, 105–111.
7 H. Rau, in *Photoreactive Organic thin films*, ed. Zouheir Sekkat, Wolfgang Knoll, Academic Press, San Diego, California (USA), 1st edition, 2002, 1, 1, 3-47.
8 Z. Li, H. Wang, M. Chu, P. Guan, Y. Zhao, Y. Zhao and J. Wang, *RSC Adv*, 2017, **7**, 44688–44695.
9 J. Eastoe and A. Vesperinas, *Soft Matter*, 2005, **1**, 338-347.
10 S. D. Pritzl, J. Morstein, N. A. Pritzl, J. Lipfert, T. Lohmüller and D. H. Trauner, *Commun mater*, 2025, **6**, 59, https://doi.org/10.1038/s43246-025-00773-8
11 R. S. H. Liu and L. U. Colmenares, *PNAS*, 2003, **25**, 14639-14644.

12 M. Cammarata, M. Levantino, F. Schotte, P. A. Anfinrud, F. Ewald, J. Choi, A. Cupane, M. Wulff and H. Ihee, *Nat Methods*, 2008, **5**, 881–886.
13 J. Woodhouse, G. Nass Kovacs, N. Coquelle, L. M. Uriarte, V. Adam, T. R. M. Barends, M. Byrdin, E. de la Mora, R. Bruce Doak, M. Feliks, M. Field, F. Fieschi, V. Guillon, S. Jakobs, Y. Joti, P. Macheboeuf, K. Motomura, K. Nass, S. Owada, C. M. Roome, C. Ruckebusch, G. Schirò, R. L. Shoeman, M. Thepaut, T. Togashi, K. Tono, M. Yabashi, M. Cammarata, L. Foucar, D. Bourgeois, M. Sliwa, J. P. Colletier, I. Schlichting and M. Weik, *Nat Commun*, 2020, **11**, 741, https://doi.org/10.1038/s41467-020-14537-0
14 Weber, L. Pithan, A. Zykov, S. Bommel, F. Carla, R. Felici, C. Knie, D. Bléger and S. Kowarik, *Journal of Physics Condensed Matter*, 2017, **29**, 434001.
15 H. Poddar, D. J. Heyes, G. Schirò, M. Weik, D. Leys and N. S. Scrutton, *The FEBS Journal*, 2022, **289**, 576-595.
16 J. Royes, V. A. Bjørnestad, G. Brun, T. Narayanan, R. Lund and C. Tribet, *J Colloid Interface Sci*, 2022, **610**, 830–841.
17 M. F. Ober, A. Müller-Deku, A. Baptist, B. Ajanović, H. Amenitsch, O. Thorn-Seshold and B. Nickel, *Nanophotonics*, 2022, **11**, 2361–2368.
18 F. Schotte, H. S. Cho, F. Dyda and P. Anfinrud, *Struct Dyn*, 2024, **11**, 021303.
19 H. S. Cho, F. Schotte, N. Dashdorj, J. Kyndt, R. Henning and P. A. Anfinrud, *J Am Chem Soc*, 2016, **138**, 8815–8823.
20 M. Levantino, G. Schirò, H. T. Lemke, G. Cottone, J. M. Glownia, D. Zhu, M. Chollet, H. Ihee, A. Cupane and M. Cammarata, *Nat Commun*, 2015, **6**, 7772.
21 L. Buglioni, F. Raymenants, A. Slattery, S. D. A. Zondag and T. Noël, *Chem Rev*, 2022, **122**, 2752-2906.
22 G. M. Whitesides, *Nature*, 2006, **44**, 368-373.
23 Y. Liu and X. Jiang, *Lab Chip*, 2017, **17**, 3960-3978.
24 A. Ghazal, J. P. Lafleur, K. Mortensen, J. P. Kutter, L. Arleth and G. V. Jensen, *Lab Chip*, 2016, **16**, 4263-4295.
25 B. F. B. Silva, *Phys Chem Chem Phys*, 2017, **19**, 26390-23703.
26 M. A. Levenstein, C. Chevallard, F. Malloggi, F. Testard and O. Taché, *Lab Chip*, 2025, **25**, 1169-1227.
27 D. Baigl, *Lab Chip*, 2012, **12**, 3637-3653.
28 D. Cojoc, H. Amenitsch, E. Ferrari, S. C. Santucci, B. Sartori, M. Rappolt, B. Marmiroli, M. Burghammer and C. Riekel, *Appl Phys Lett*, 2010, **97**, 244101.
29 S. C. Santucci, D. Cojoc, H. Amenitsch, B. Marmiroli, B. Sartori, M. Burghammer, S. Schoeder, E. Dicola, M. Reynolds and C. Riekel, *Anal Chem*, 2011, **83**, 12, 4863-4870.
30 E. Vergucht, T. Brans, F. Beunis, J. Garrevoet, S. Bauters, M. De Rijcke, D. Deruytter, C. Janssen, C. Riekel, M. Burghammer and L. Vincze, *J Synchrotron Radiat*, 2015, **22**, 1096–1105.
31 R. K. Ramamoorthy, E. Yildirim, I. Rodriguez-Ruiz, P. Roblin, L. M. Lacroix, A. Diaz, R. Parmar, S. Teychené and G. Viau, *Lab Chip*, 2023, **24**, 327–338.
32 Y. Cao, J. Floehr, S. Ingebrandt and U. Schnakenberg, *Micromachines*, 2021, **12**, 632.
33 A. El Hasni, S. Pfirrmann, A. Kolander, E. Yacoub-George, M. König, C. Landesberger, A. Voigt, G. Grützner and U. Schnakenberg, *Microfluid Nanofluidics*, 2017, **21**,41.
34 M. M. Roos, A. Winkler, M. Nilsen, S. B. Menzel and S. Strehle, *International Journal of Precision Engineering and Manufacturing - Green Technology*, 2022, **9**, 43–57.
35 M. Rödl, K. Küssner and H. A. Schwartz, *Z Anorg Allg Chem*, 2024, **650**, e202400018.
36 H. Amenitsch, M. Rappolt, M. Kriechbaum, H. Mio, P. Laggner and S. Bernstorff, *J. Synchrotron Rad*, 1998, **5**, 506–508.
37 M. Burian, C. Meisenbichler, D. Naumenko and H. Amenitsch, *J Appl Crystallogr*, 2022, **55**, 677–685.
38 O. Glatter, *Scattering Methods and their Application in Colloid and Interface Science*, Elsevier, Amsterdam, Netherlands 2018.
39 J. Trewhella, A. P. Duff, D. Durand, F. Gabel, J. M. Guss, W. A. Hendrickson, G. L. Hura, D. A. Jacques, N. M. Kirby, A. H. Kwan, J. Pérez, L. Pollack, T. M. Ryan, A. Sali, D. Schneidman-Duhovny, T. Schwede, D. I. Svergun, M. Sugiyama, J. A. Tainer, P. Vachette, J. Westbrook and A. E. Whitten*, Acta Crystallogr D Struct Biol*, 2017, **73**, 710–728.
40 D. I. Svergun, *J Appl Crystallogr*, 1992, **25**, 495–503.
41 M. Burian, B. Marmiroli, A. Radeticchio, C. Morello, D. Naumenko, G. Biasiol and H. Amenitsch, *J Synchrotron Radiat*, 2020, **27**, 51-59.
42 H. S. Cho, F. Schotte, V. Stadnytskyi, A. Dichiara, R. Henning and P. Anfinrud, *Journal of Physical Chemistry B*, 2018, **122**, 11488–11496.
43 R. Haider, B. Marmiroli, I. Gavalas, M. Wolf, M. Matteucci, R. Taboryski, A. Boisen, E. Stratakis and H. Amenitsch, *Microelectron Eng*, 2018, **195**, 7–12.
44 D. Orthaber, A. Bergmann and O. Glatter*, J. Appl. Cryst*, 1999, **33**, 218-225.
45 W. A. Eaton, E. R. Henry, J. Hofrichter, S. Bettati, C. Viappiani and A. Mozzarelli*, IUBMB Life*, 2007, **59** (8-9), 586-599.
46 J. Lal, M. Maccarini, P. Fouquet, N. T. Ho, C. Ho and L. Makowski, *Protein Science*, 2017, **26**, 505–514.
47 R. Haider, B. Sartori, A. Radeticchio, M. Wolf, S. D. Zilio, B. Marmiroli and H. Amenitsch, *J Appl Crystallogr*, 2021, **54**, 132–141.
48 R. Prassl, M. Pregetter, H. Amenitsch, M. Kriechbaum, R. Schwarzenbacher, J. M. Chapman and P. Laggner, *PLoS One*, 2008, **3**, 12, e4079.

# Supplementary Information

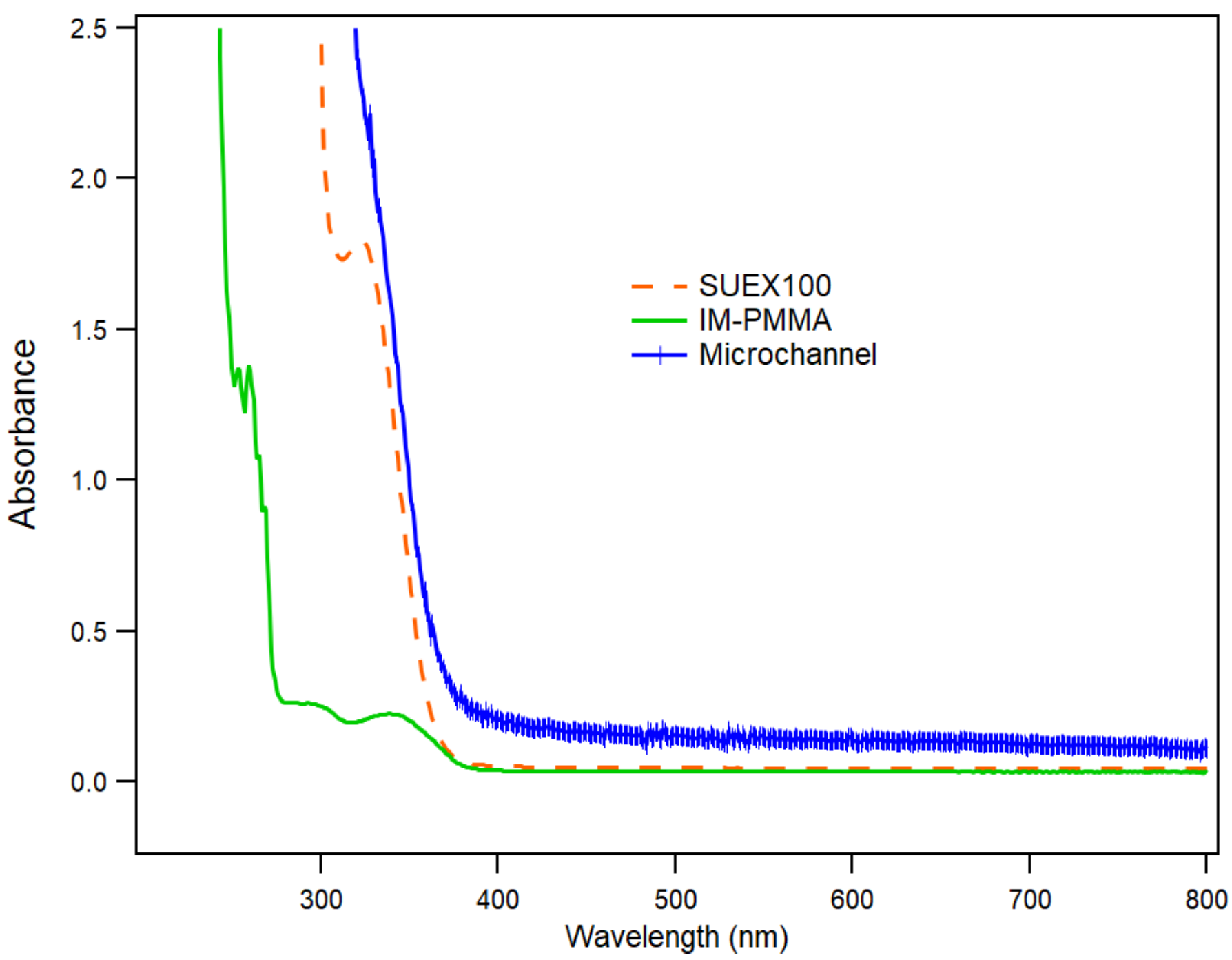

**Figure S1**. UV-Vis absorbance spectra of the materials employed for the fabrication of the microfluidic device, as well as spectrum of the empty channel itself. IM-PMMA has been used as top cover of the channel, as it is almost transparent to UV light down to a wavelength of 280 nm (continuous green line). The SUEX 100 slide (orange dashed line), and therefore the microchannel (blue line with vertical bars), are only transparent to visible light.

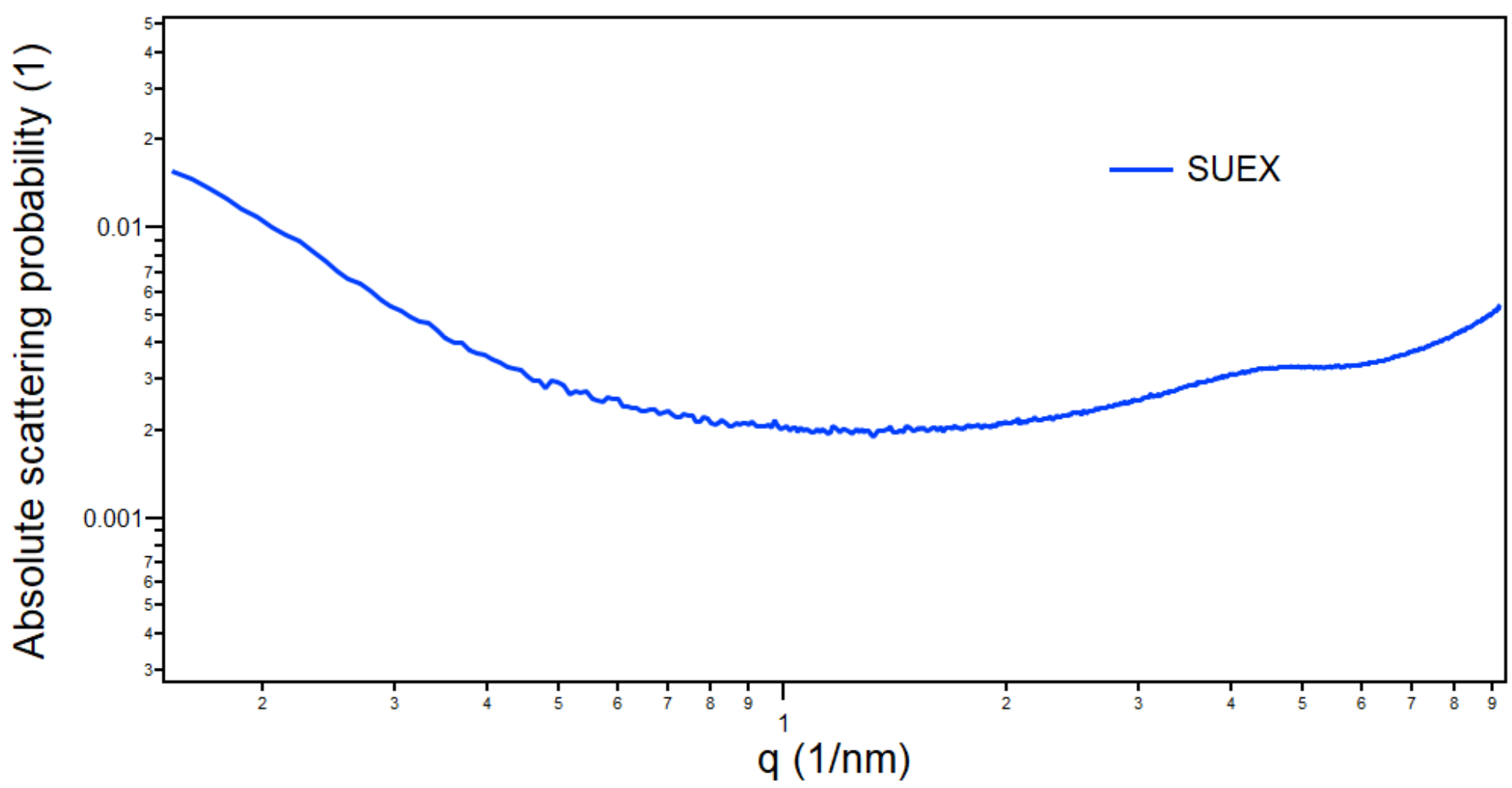

**Figure S2.** Absolute scattering probability of SUEX dry film resist.

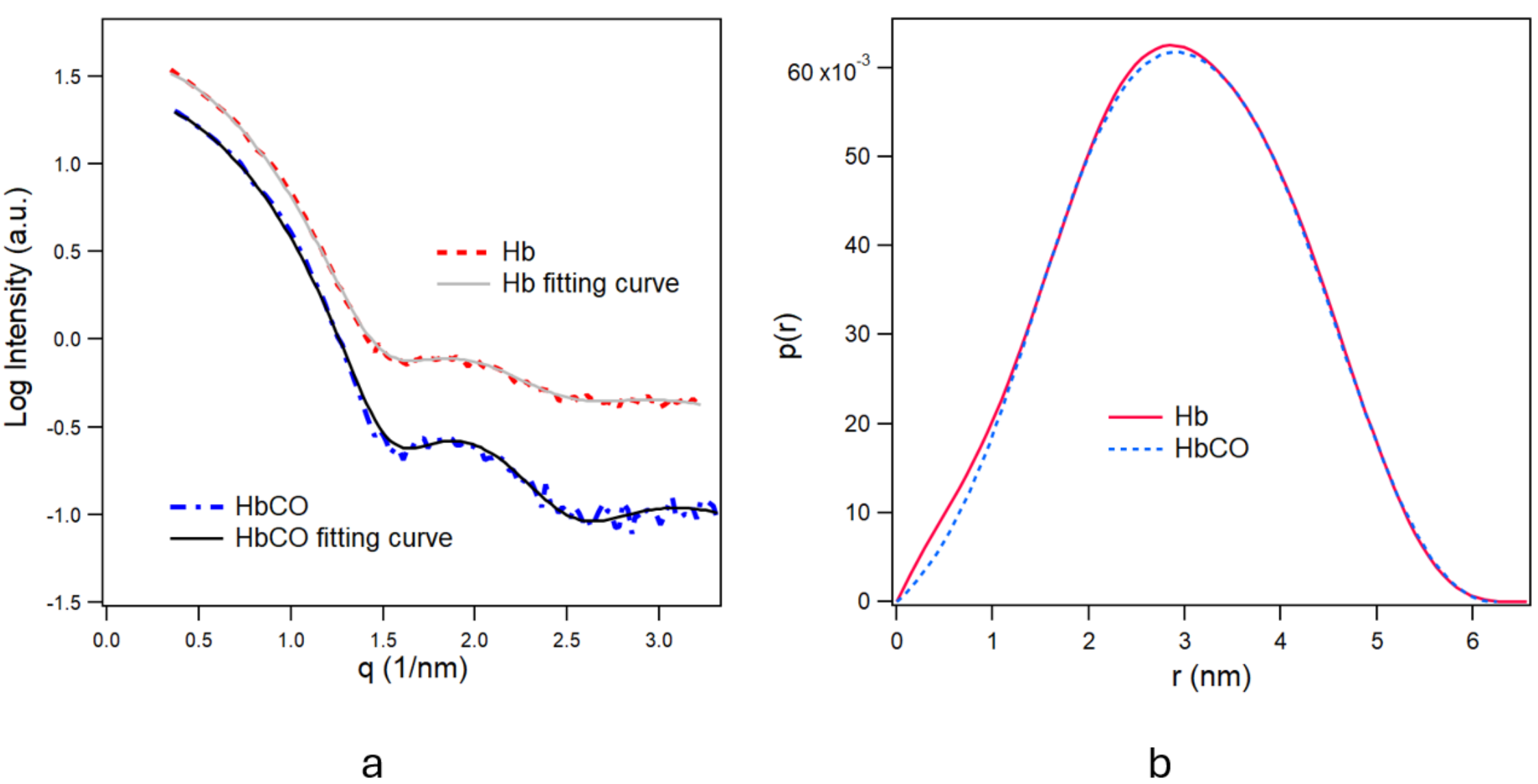

**Figure S3**. (a) Scattering curves and corresponding fitting curves calculated using GNOM program: Hb (red dashed line) and fitting (continuous grey line); HbCO (blue dashed with dots line) and fitting (continuous black line). Curves have been shifted vertically for clarity. (b) Pair Distance Distribution Function of Hb (red continuous line) and of HbCO (blue dashed line).

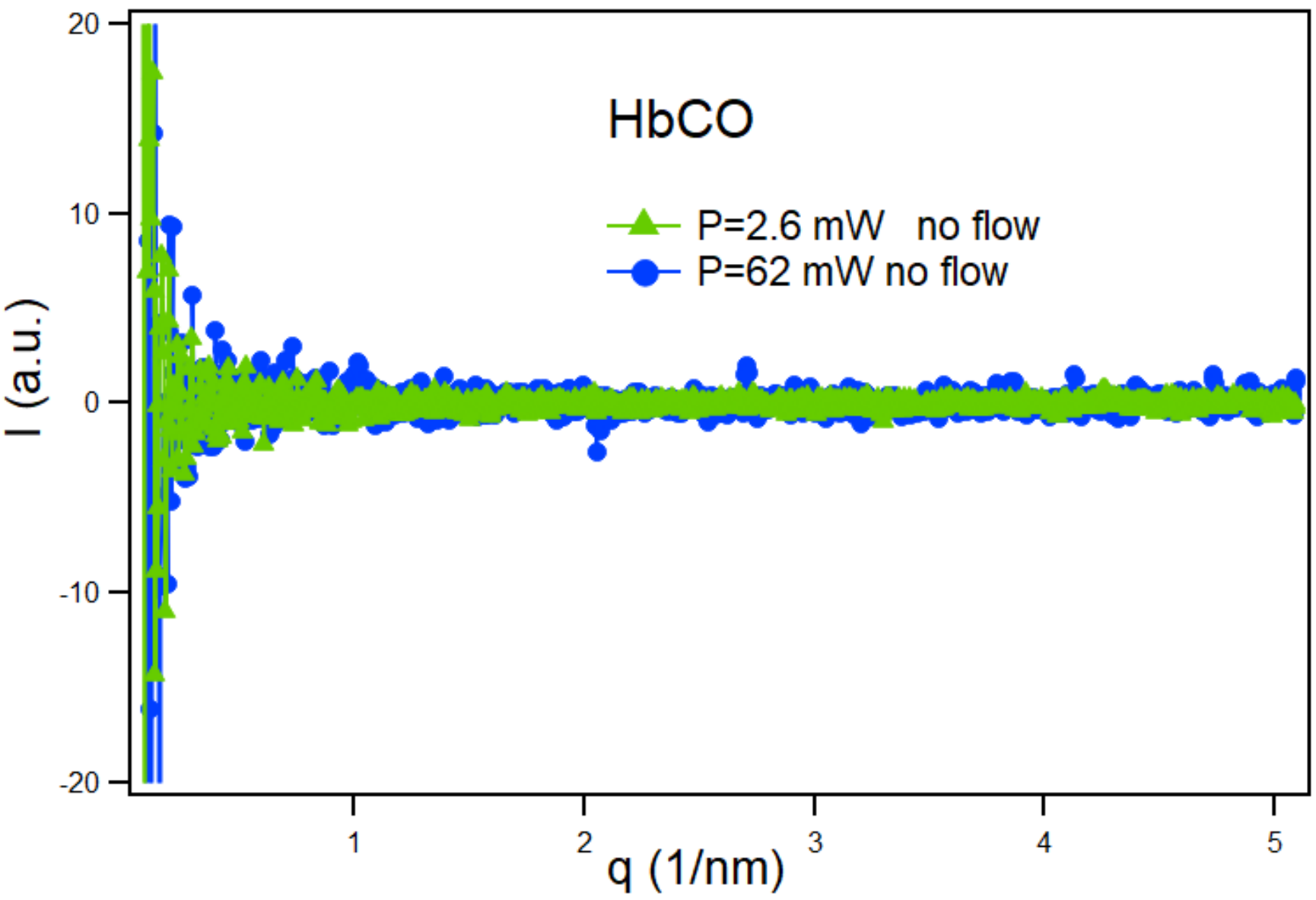

**Figure S4**. Scattering curve of HbCO with laser ON subtracted by the one with laser OFF. Two different transmission values of the laser have been used: T 5% corresponding to P= 2.6 mW (green curve with triangles), T 50% corresponding to P= 61.7 mW (blue curve with circular dots) with no liquid flow.

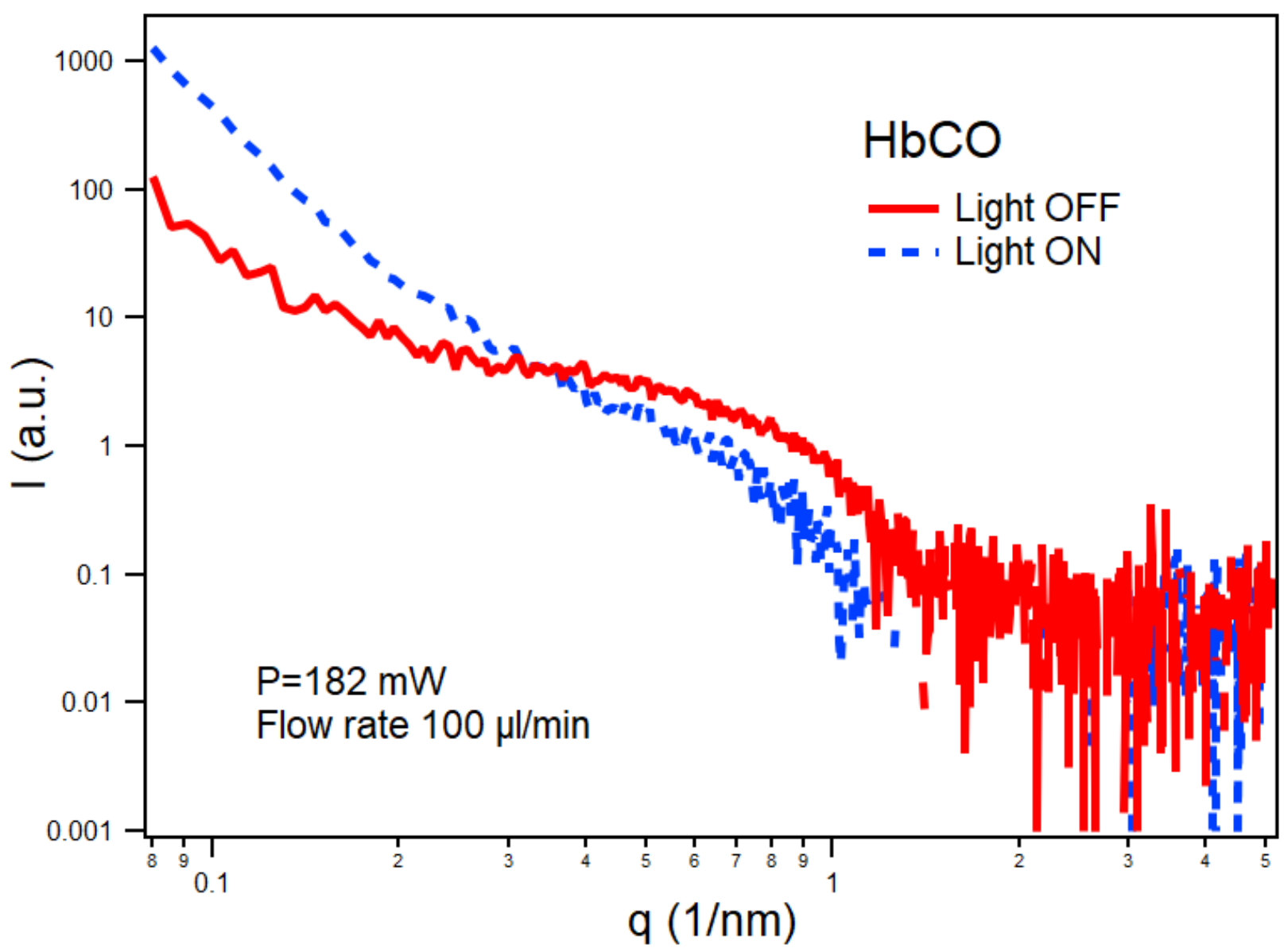

**Figure S5.** Scattering curve of HbCO in the microchannel with an exposure time of 10 s while pumping with a flow rate of 100 µl/min. The red continuous curve corresponds to the protein before irradiation, the blue dashed one to the same protein exposed to laser light.

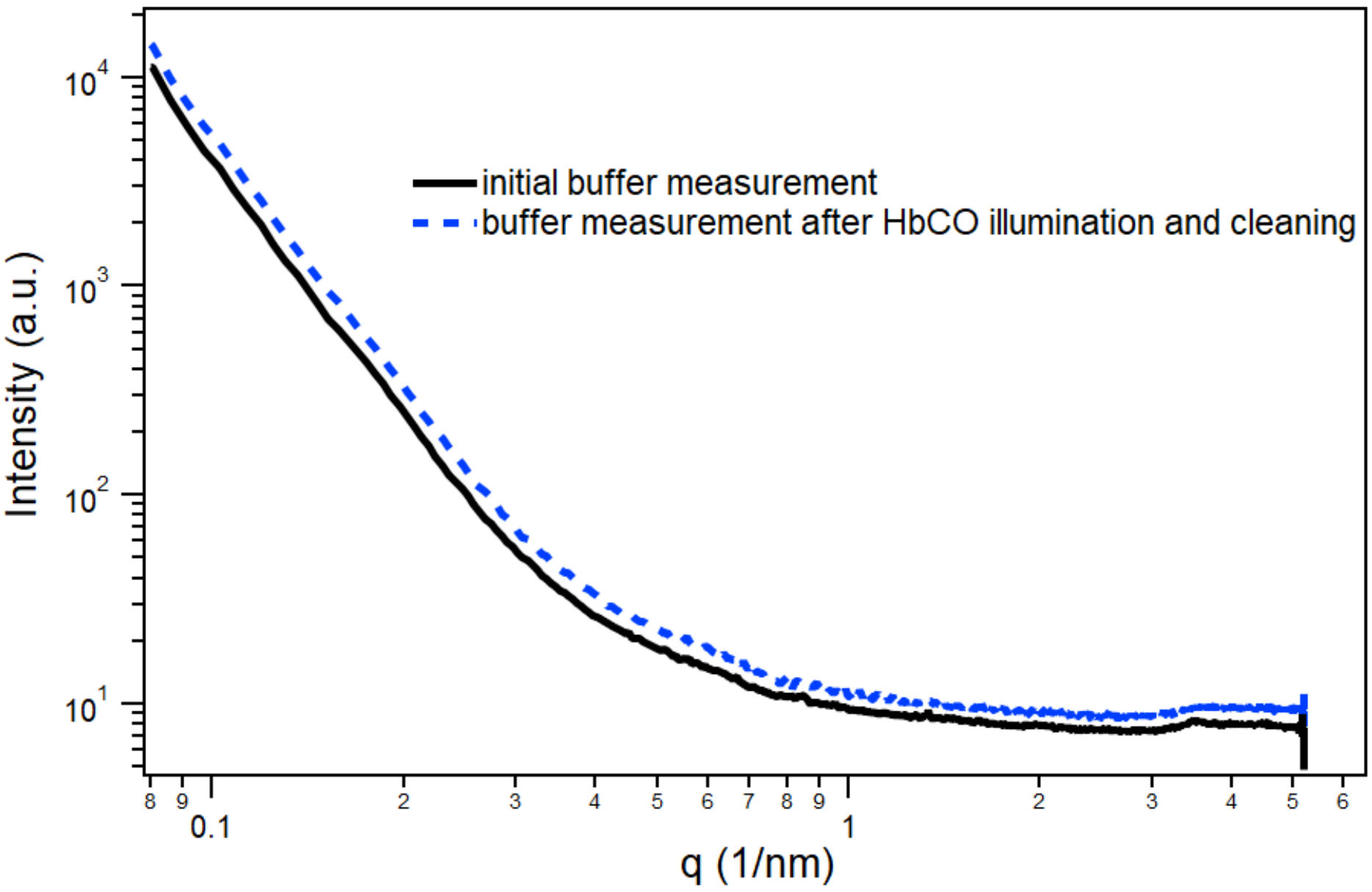

**Figure S6**. Scattering curves of the microchannel containing buffer before and after cleaning. An initial buffer measurement was taken for 10 s, then HbCO protein was measured taking 1000 images of 0.1 s during laser irradiation with P= 182 mW. Afterwards, the device was cleaned with a 1% v/v aqueous solution of Hellmanex III, followed by rinsing with Milli-Q water. The microchannel was then filled with buffer and measured for 10 s. The curves are vertically shifted for clarity.

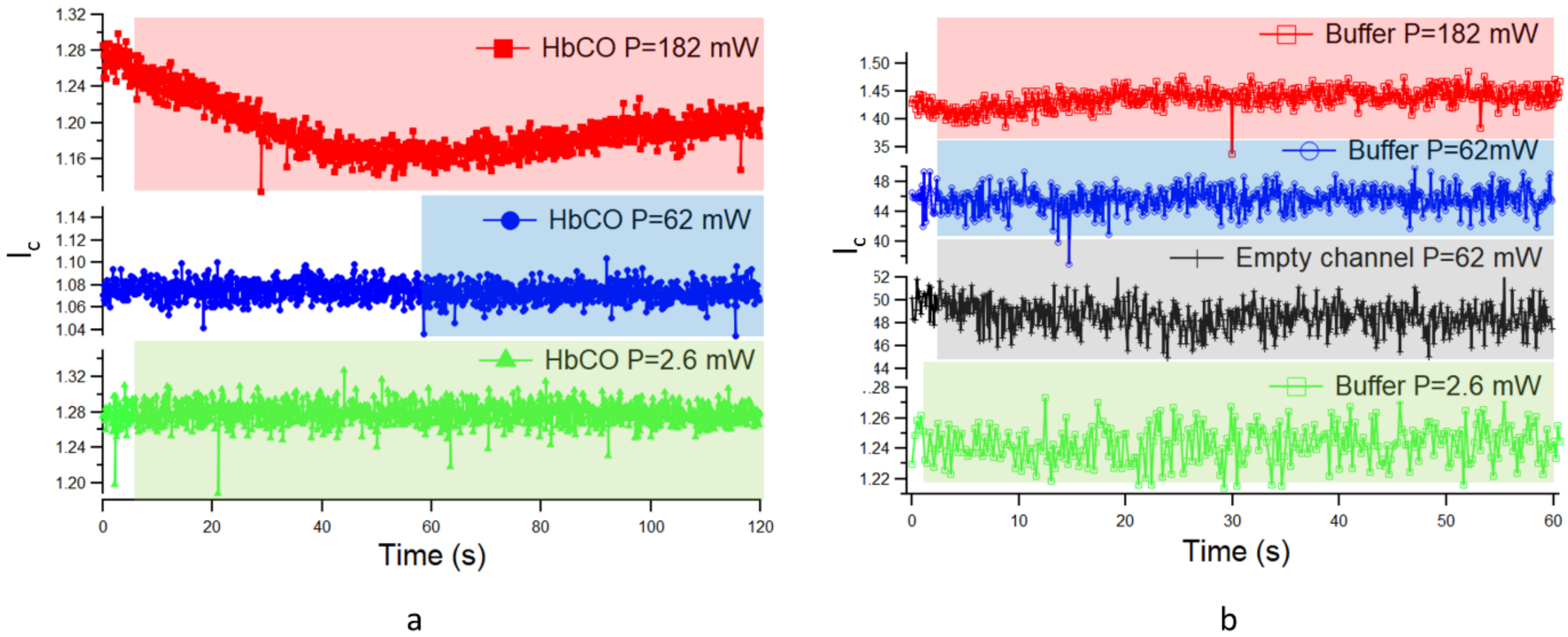

**Figure S7. (a)** Calculated correlation length vs SAXS exposure time for the protein inside the microfluidic device flowing at 100 µl/min. During the SAXS measurement the light is turned ON. The shadowed area corresponds to the measurements done with laser ON. Different laser power values are shown. The laser with P=62 mW (blue shadow) was turned on around one minute after starting the measurement. **(b)** Calculated correlation length vs SAXS exposure time for the buffer inside the microfluidic device flowing at 100 µl/min for different laser power values and for the empty channel.

## 1. SAXS data manipulation

In the following, the formulas employed to calculate relevant SAXS parameters for the present study are reported:[1]

*Radius of gyration $R_g$ :*

$$I(q) = I(0)e^{-\frac{q^2 Rg^2}{3}} \qquad (1)$$

q scattering vector

I(q) Scattering intensity

I(0) Forward scattering intensity

*Pair Distance Distribution Function (PDDF) p(r)*

$$p(r) = \frac{1}{2\pi^2}\int_0^\infty I(q)qrsin\,(qr)dq \qquad (2)$$

*Invariant Q:*

$$Q = \int_0^\infty I(q)q^2 dq$$

*Correlation length $l_c$*

$$lc = \pi \frac{\int_0^{\infty} I(q)qdq}{Q}$$

As in the present case it was not possible to integrate in the range from q= 0 to q= ∞, we integrated in the measured q range, therefore:

$$Q = \int_{q_{min}}^{q_{max}} I(q)q^2 dq \qquad (3)$$

$$lc = \pi \frac{\int_{q_{min}}^{q_{max}} I(q)qdq}{Q} \qquad (4)$$

Reference

1 O. Glatter, *Scattering Methods and their Application in Colloid and Interface Science*, Elsevier, Amsterdam, Netherlands 2018.

## 2. Estimate of SUEX attenuation length

According to the method commonly employed in X-ray lithography[2], the first step to estimate the transparency of a material to X-rays of a given wavelength is to determine its attenuation length, as it is independent from the material thickness. For this, the chemical composition of the material must be at least approximately known.
In the present case, SUEX is composed of a mixture of epoxy resins. The major components are Bisphenol A (BPA) and Bisphenol A diglycidyl ether (DGEBA). BPA can be described as $C_{15}H_{16}O_2$ with a density of 1.2 g/cm$^3$, DGEBA is $C_{21}H_{24}O_4$ with density of 1.17 g/cm$^3$.
In order to estimate SUEX transparency to 8 KeV X-rays, we therefore determined the attenuation length of these two components.
X-ray wavelength, chemical formula and density were inserted in the calculation sheet prepared by the Center for X-ray Optics (CXRO) of Lawrence Berkeley National Laboratory's (LBNL): https://henke.lbl.gov/optical_constants/
The resulting attenuation lengths are 1.642 mm for BPA, and 1.587 for DGEBA. The attenuation length of the mixture is expected to be an intermediate between these two values.
Transmission can also be calculated in the same website, considering the thickness of the material. For SUEX of 0.5 mm the value of transmission is similar for the two epoxy resins and approximately equal to 73%.

Reference

2 M. Tormen, G. Grenci, B. Marmiroli, F. Romanato in *Nanolithography*, ed. Stefan Landis, ISTE Ltd. and John Wiley & Sons, Inc., London, 2011, 1,1, 1-79.